\documentclass{PoS}
\pdfoutput=1

\title{A comparative study of two lattice approaches to two-body systems}

\ShortTitle{A comparative study of two lattice approaches to two-body systems}

\author{\speaker{Bruno Charron} \\
        Physics Dept., University of Tokyo, 7-3-1 Hongo, Bunkyo, Tokyo 113-0033, Japan\\
        E-mail: \email{charron@riken.jp}}

\author{for HAL QCD Collaboration}

\abstract{
  We present a method to extract the spectrum of two-particle systems on the lattice from wave functions computed in lattice simulations.
  The energies of the Hamiltonian eigenstates are extracted from the eigenvalues of a matrix, similar to a potential, constrained by the wave functions.
  This method is compared with the traditional variational method in the isospin 2 $\pi\pi$ system.
}

\FullConference{31st International Symposium on Lattice Field Theory LATTICE 2013\\
                   July 29 -- August 3, 2013\\
                   Mainz, Germany}

\usepackage{amsmath}
\usepackage{amssymb}

\usepackage{caption}
\usepackage{subcaption}

\begin{document}

\section{Introduction}

Since its introduction by M.~L\"uscher \cite{luescher_formula}, the finite size formula relating the finite volume spectrum of a two-particle system to its infinite volume phase shifts has allowed the systematic study of scattering states, bound states, resonances and other interesting QCD phenomena.
In order to extract the finite volume spectrum from the lattice, a variational technique \cite{luescher_var} is usually used to disentangle the mixing of a set of states created from lattice operators with the QCD eigenstates.

Recently, an alternative method has been introduced \cite{hal_first} to study two-particle systems from wave functions computed on the lattice.
These wave functions can be shown to satisfy a Schr\"odinger-like equation with a non-local but energy-independent potential.
Once the potential is obtained from lattice data, one can solve the equation in infinite volume to compute the phase shifts of the system.

In this paper, we present a method to obtain the finite volume spectrum which combines the two previous ideas and compare it to the traditional variational approach.
The second section contains the formulation of this method, after a review of the variational method.
In the third section, we present the numerical setup used for the comparison of the two methods.
Finally we show the results of the comparison and conclude.

\section{Formulation}

\subsection{Variational method}

Let $\{\mathcal{S}_j\}_{j=1}^{N}$ and $\{\mathcal{O}_i\}_{i=1}^{N}$ be two sets of operators annihilating two-particle states on the same time slice.
We define the $N \times N$ correlation matrices with elements
\begin{align}
  \label{eq:corr_func}
  \textstyle
  C_{ij}(t) \equiv \langle \mathcal{O}_i(t) \mathcal{S}_j^\dagger(t_S) \rangle_L = \sum_n \langle 0 | \mathcal{O}_i | n\rangle e^{-W_n(t-t_s)} \langle n | \mathcal{S}_j^\dagger | 0\rangle,
\end{align}
where $\langle \cdot \rangle_L$ denotes the expectation value in lattice QCD of the associated field functionals, $|0 \rangle$ is the vacuum state and $| n \rangle$, for $n \geq 1$, are eigenstates of the Hamiltonian $\hat H$ with increasing energies $W_n$.
In this sum and later, we only consider the eigenstates $|n \rangle$ which overlap with at least one $\mathcal{S}_j^\dagger | 0\rangle$ source state.

Define the $N \times N$ matrices $P$ and $Q$, with elements $P_{ni} = \langle n | \mathcal{O}_i^\dagger | 0\rangle$ and $Q_{nj} = \langle n | \mathcal{S}_j^\dagger | 0\rangle$, representing the mixing of the sink and source states with the first $N$ eigenstates of the Hamiltonian.
Eq.~\eqref{eq:corr_func} can then be written in the matrix form
\begin{align}
  \label{eq:corr_func_mat}
  \textstyle
  C(t) = P^\dagger \, D(t - t_S) \, Q + \mathcal{O}(e^{-W_{N+1} (t-t_s)}),
\end{align}
where the matrix $D$ is diagonal with elements $D_{nn}(\Delta t) = e^{-W_n \Delta t}$ and the second term contains the contributions of all eigenstates of the Hamiltonian with energies higher than $W_N$.

Assuming that the matrix $Q$ is invertible, i.e. that the projections of the first $N$ eigenstates of the Hamiltonian on the vector space spanned by the source states $\mathcal{S}_j^\dagger | 0\rangle$ are linearly independent, we have
$C(t + \Delta t)Q^{-1} = C(t) \, Q^{-1} \, D(\Delta t) + \mathcal{O}(e^{-W_{N+1} (t-t_s)})$.
This means that the solutions of the generalized eigenvalue problem (GEVP)
\begin{align}
  \label{eq:gevp}
  C(t + \Delta t) \; v_n(t, \Delta t) = \lambda_n(t, \Delta t) \; C(t) \; v_n(t, \Delta t)
\end{align}
are such that $v_n(t, \Delta t)$ converges to the $n$-th column of $Q^{-1}$ and $\lambda_n(t, \Delta t)$ to $e^{-W_n\Delta t}$ when $|t - t_S| \to \infty$.
It is thus possible to use the following estimators
\begin{align}
  \label{eq:gevp_eeff}
  \textstyle
  W_n^\textrm{eff}(t) = \frac{1}{a} \log \frac{\lambda_n(t, \Delta t)}{\lambda_n(t, \Delta t + a)},
\end{align}
with $a$ the (isotropic) lattice spacing, to obtain the energies of the first $N$ Hamiltonian eigenstates as $W_n = \lim_{|t-t_S| \to \infty} W_n^\textrm{eff}(t)$.
Such a choice of estimator, for fixed $\Delta t$, was proven \cite{blossier} more efficient than taking a fixed reference time for the GEVP.

\subsection{Potential method}

Lattice wave functions can be seen as correlation matrices with $N$ source operators but $N_s^3 \gg N$ sink operators, where $N_s$ is the number of lattice points in each spatial direction.
Specifically, the sink operators $\mathcal{O}_i$ are taken to be interpolators for two point-like particles on the same time slice and with a spatial separation $\textbf{r}_i$.
Then, Eq.~\eqref{eq:corr_func} can be used to define $N_s^3 \times N$ matrices $C(t)$.

Introducing the following, more convenient, notations,
\begin{align*}
  \textstyle
  \psi_j(\textbf{r}_i, t) = C_{ij}(t) e^{2mt}, \quad \phi_n(\textbf{r}_i) = \langle 0 | \mathcal{O}_i | n \rangle, \quad \Delta W_n = W_n - 2m,
\end{align*}
Eq.~\eqref{eq:corr_func} is simply written as $\psi_j(\textbf{r}, t) = \sum_n Q_{nj}e^{-\Delta W_n(t-t_S)} \phi_n(\textbf{r})$.
For simplicity, assume that there is some intermediate time $t_0$ such that only the first $N_\textrm{eff}$ eigenstates of the Hamiltonian (possibly $N_\textrm{eff} \gg N$) effectively contribute to the wave functions $\psi_j(\textbf{r}, t)$ with $t \geq t_0$.

While the elements $Q_{nj}$ are unknown, we can use the known time-dependence to find
\begin{align}
  \label{eq:time_dep_a}
  \textstyle
  \forall j \leq N, \; t \geq t_0, \quad \left(\frac{\nabla^2}{m} - \partial_t + \frac{1}{4m} \partial_t^2 \right) \psi_j(\textbf{r}, t)
  = \sum_{n\leq N_\textrm{eff}} Q_{nj}e^{-\Delta W_n(t-t_S)} \left(\frac{\nabla^2}{m} + \frac{k_n^2}{m} \right) \phi_n(\textbf{r})
\end{align}
where $\nabla^2$ and $\partial_t$ are the discretized Laplacian and time derivative operator respectively.
We have used the identity $\Delta W_n + \frac{(\Delta W_n)^2}{4m} = \frac{k_n^2}{m}$ with $W_n = 2 \sqrt{k_n^2 + m^2}$ and
discretization errors from the time derivative are neglected.

We have seen that the wave functions $\psi_j(\cdot, t)$ are linear combinations of the wave functions $\phi_n$ with source- (index $j$) and time- (index $t$) dependent coefficients.
For this reason, we consider pairs of indices $c=(j,t)$ to label the wave functions $\Psi_c(\textbf{r}) = \psi_j(\textbf{r}, t)$.

Consider the wave functions $\Psi_c$ and $\phi_n$ as $N_s^3$-dimensional vectors with components $\psi_j(\textbf{r}, t)$ and $\phi_n(\textbf{r})$ respectively.
Then, if an $N_s^3 \times N_s^3$ matrix $U$ satisfies
\begin{align}
  \label{eq:def_u}
  \textstyle
  \forall c \in \mathcal{C}, \quad U \cdot \Psi_c =  \left(\frac{\nabla^2}{m} - \partial_t + \frac{1}{4m} \partial_t^2 \right) \Psi_c
\end{align}
for some set $\mathcal{C}$ of pairs of source and time indices (all larger than $t_0$), Eq.~\eqref{eq:time_dep_a} leads to
\begin{align}
  \label{eq:contraction_zero}
  \textstyle
  \forall c=(j,t) \in \mathcal{C}, \quad \sum_{n\leq N_\textrm{eff}} Q_{nj}e^{-\Delta W_n(t-t_S)} \left[ U \cdot \phi_n - \left(\frac{\nabla^2}{m} + \frac{k_n^2}{m}\right) \phi_n \right] = 0.
\end{align}

The consequence is the following.
Construct an $N_\textrm{eff} \times |\mathcal{C}|$ matrix $\tilde Q$ ($|\mathcal{C}|$ being the size of $\mathcal{C}$) with components $\tilde Q_{nc} = Q_{nj}e^{-\Delta W_n(t-t_S)}$ for $c=(j,t) \in \mathcal{C}$.
If the matrix $\tilde Q$ has full row rank, one finds that
\begin{align}
  \label{eq:eigv_phi}
  \textstyle
  \forall n \leq N_\textrm{eff}, \quad U \cdot \phi_n - \left(\frac{\nabla^2}{m} + \frac{k_n^2}{m}\right) \phi_n = 0.
\end{align}

Equivalently, $\phi_n$ for $n\leq N_\textrm{eff}$ are eigenvectors of the matrix $H = -\frac{\nabla^2}{m} + U$ with respective eigenvalues $\frac{k_n^2}{m}$.
Therefore, the energies of the first $N_\textrm{eff}$ eigenstates of the Hamiltonian can be recovered from the spectrum of $H$.
By similarity with the Schr\"odinger equation, the operator $U$ is called a potential.

\subsection{Qualitative comparison}

The idea behind the variational method is to recover the contribution of each eigenstate of the Hamiltonian by inverting the mixing $Q$ using the known behavior of the time evolution.
This inversion is achieved by solving a GEVP, Eq.~\eqref{eq:gevp}.

In the potential method, information on the eigenstates of the Hamiltonian is recovered indirectly by building a linear operator such that the linear combinations $\Psi_c$ satisfy an equation, \eqref{eq:def_u}, which reduces to an eigenvalue problem for the underlying wave functions $\phi_n$, \eqref{eq:eigv_phi}.

It is to be noted that in the variational method framework, the number of energies extracted is at most equal to the number of sources.
In the potential method, the use of pairs of source and time indices effectively expands the number of sources.
It is equivalent to considering several sources, constructed from the same operator but shifted in time, e.g. $S_j^\dagger|0\rangle$ and $e^{-\hat H \Delta t}S_j^\dagger|0\rangle$.
This trick can also be used in the variational method approach but usually leads to poor results.
In the potential method, the use of a large number of sink operators can allow a better separation of these similar source operators.

We now review the assumptions of the two methods.
For the variational method, the $N \times N$ matrix $Q$ is assumed invertible.
For the potential method, the $N_\textrm{eff} \times |\mathcal{C}|$ matrix $\tilde Q$ is assumed to have full row rank.
Since $Q$ is unknown, these conditions can only be assumed but one can get an idea of their validity by looking at the data.

Furthermore, in the potential method, we assume that a matrix $U$ satisfying Eq.~\eqref{eq:def_u} exists, which requires the wave functions $\Psi_c$ to be linearly independent.
This is a weak condition and easily checked, as opposed to a condition on $Q$.

The derivation of the potential method could also be carried out by chosing to define directly a matrix $H$ such that $\forall c \in \mathcal{C}, \;\; H \cdot \Psi_c =  \left(- \partial_t + \frac{1}{4m} \partial_t^2 \right) \Psi_c$ instead of Eq.~\eqref{eq:def_u}.
The difficulty is then to chose a specific form for the matrix $H$.
As will be detailed in a future paper \cite{future_paper}, the formulation in terms of a potential is found in practice more convenient.

\section{Numerical setup for the comparison}

In the previous section, we have presented the variational method and the potential method which both allow to extract the energies of the first few eigenstates of the Hamiltonian.
We have compared these two methods in the isospin $I=2$ channel of the two-pion system.

Consider the following source and sink interpolators:

$\overline{\pi\pi}(\textbf{q}) = [\bar u(\textbf{q}) \gamma_5 d(\textbf{0})] [\bar u(-\textbf{q}) \gamma_5 d(\textbf{0})]$ where $\bar u, d(\textbf{q})$ are momentum wall quarks,

$\pi\pi(\textbf{r}, t) = P^{(A_1^+)} \cdot \sum_\textbf{x} \pi^+(\textbf{x} + \textbf{r}, t) \pi^+(\textbf{x}, t)$ with $\pi^+(\textbf{x})$ a local interpolator for the pion and $P^{(A_1^+)}$ the projection on the $A_1^+$ representation of the cubic group.

\begin{table}
\centering
\renewcommand{\arraystretch}{1.2}
\begin{tabular}{ c | c | c | }
  \cline{2-3}
  & Variational method & Potential method \\  \hline
  \multicolumn{1}{ |c| }{Source operators} & $\overline{\pi\pi}(\textbf{q})$ for $\textbf{q} \in \mathcal{Q}$ & $\overline{\pi\pi}(\textbf{q})$ for $\textbf{q}=0$ \\ \hline
  \multicolumn{1}{ |c| }{Sink operators} & $\sum_\textbf{r} e^{-i\textbf{q} \cdot \textbf{r}} \pi\pi(\textbf{r}, t)$ for $\textbf{q} \in \mathcal{Q}$ & $\pi\pi(\textbf{r}, t)$ for all $\textbf{r}$ \\ \hline
\end{tabular}
\caption{Choice of source and sink operators for the two methods. See text for details.}\label{tab:i2_ops}
\end{table}

The operators used for each method are summarized in Table~\ref{tab:i2_ops}.
For the variational method, $N = 4$ and the momenta are chosen in $\mathcal{Q} = \{(0,0,0)$, $(0,0,1)$, $(0,1,1)$, $(1,1,1)\}$ in units of $\frac{2\pi}{L}$ where $L$ is the spatial extent of the lattice.
For the potential method, $N=1$.

The computations are based on $N_f=2+1$ QCD gauge configurations generated by the PACS-CS collaboration \cite{configs_pacs} on a $32^3\times 64$ lattice with the Iwasaki gauge action at $\beta = 1.9$ and clover fermions.
The lattice spacing is $a = 0.09$ fm and the sea quark hopping parameters are $\kappa_{ud} = 0.1370$ and $\kappa_{s} = 0.1364$, making for a pion mass $m_\pi = 0.7$ GeV.

For both methods, the signal is improved by including a factor $e^{-2mt} / C_2(t)^2$ in the 4-point correlators, where $C_2(t)$ is the 2-point correlator for a time separation of $t-t_S$. At large time indices $t$, this factor converges to unity but at earlier times, it can cancel the contaminations of some high-energy eigenstates.

\section{Results}
\setlength{\abovecaptionskip}{-1ex} 

For the variational method, the energies of the first $N=4$ eigenstates of the Hamiltonian were estimated using Eq.~\eqref{eq:gevp_eeff} with a fixed $\Delta t = 3$ (times will always be given implicitely in units of $a$).

\begin{figure}
  \centering
  \includegraphics{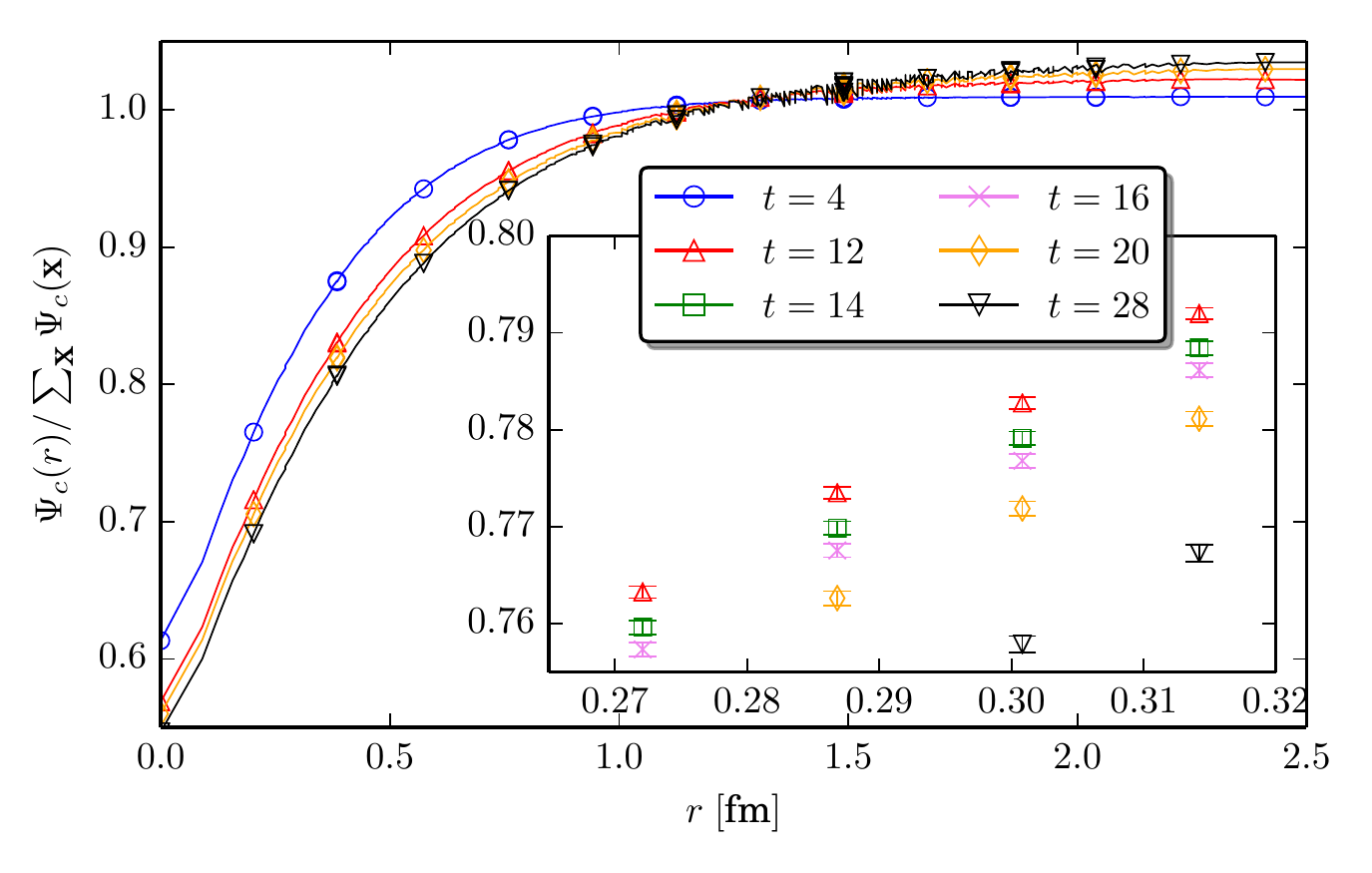}
  \caption{Time dependence of the wave function $\Psi_c$.}\label{fig:comp_wf}
\end{figure}

For the potential method, we first look at the wave functions $\Psi_c$ in Fig.~\ref{fig:comp_wf} for $c=(1, t)$ since we only use one source $j=1$.
They are normalized so as to clearly see the time-dependence of their shape.
We can see up to $t=28$ a clear time-dependence, larger than statistical errors.
This means that at least two eigenstates of the Hamiltonian contribute effectively ($N_\textrm{eff} \geq 2$) on this time range.
Since the eigenstate energies should not be too far from their free values and these are such that $W^\textrm{free}_3 - W^\textrm{free}_2 \simeq W^\textrm{free}_2 - W^\textrm{free}_1$, we deduce that at intermediate times $t \simeq 14$ we have $N_\textrm{eff} \geq 3$.
We notice that at small time indices, the tails of the wave functions are flat.
This is due to the choice of the source operator, a wall source, and the wave functions gradually converge toward $\phi_1$ at large $t$.

\begin{figure}
  \centering
  \includegraphics{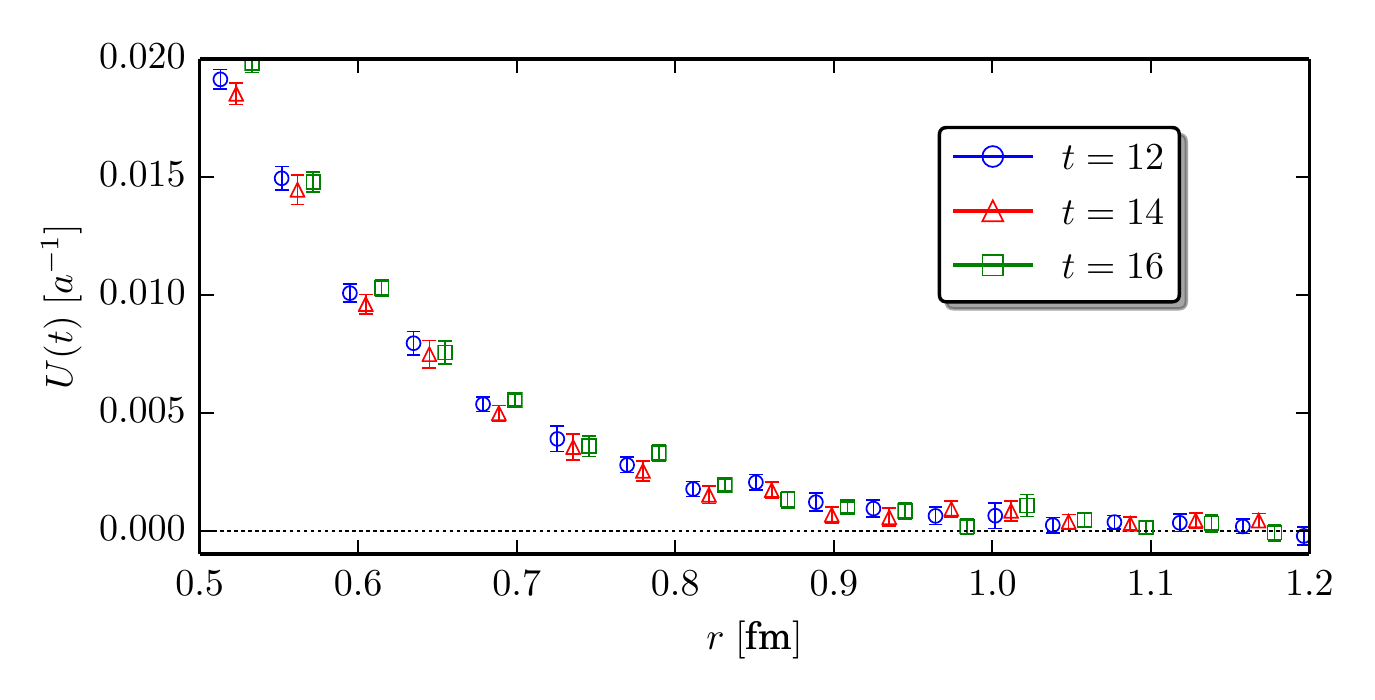}
  \caption{Time dependence of the operator $U(t)$.}\label{fig:comp_pot}
\end{figure}

For each time slice, we define a diagonal operator $U(t)$ of which the diagonal elements are set to satisfy Eq.~\eqref{eq:def_u} (with $\mathcal{C} = \{(1,t)\}$).
From Fig.~\ref{fig:comp_pot}, we can see that for $t$ ranging from $12$ to $16$, this operator is constant within statistical errors.
This means that, up to statistical errors, $U$ taken as $U(t=14)$ satisfies Eq.~\eqref{eq:def_u} for $\mathcal{C} = \{(1,12), (1,14), (1,16)\}$.
In comparison, we can see in the inset of Fig.~\ref{fig:comp_wf} that the wave functions at these time indices are clearly separated.

\begin{figure}
        \centering
        \hspace*{-.3in}
        \begin{subfigure}[b]{0.47\textwidth}
                \includegraphics{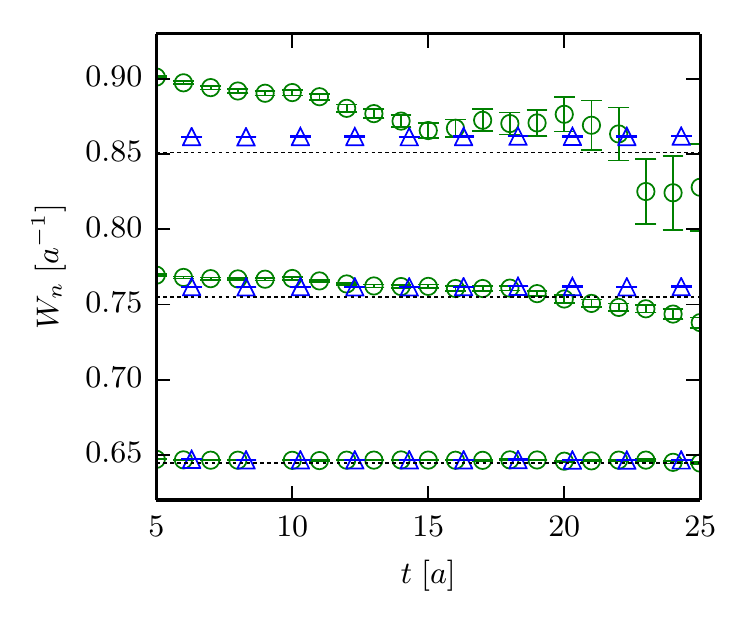}
        \end{subfigure}%
        \quad
        \begin{subfigure}[b]{0.47\textwidth}
                \includegraphics{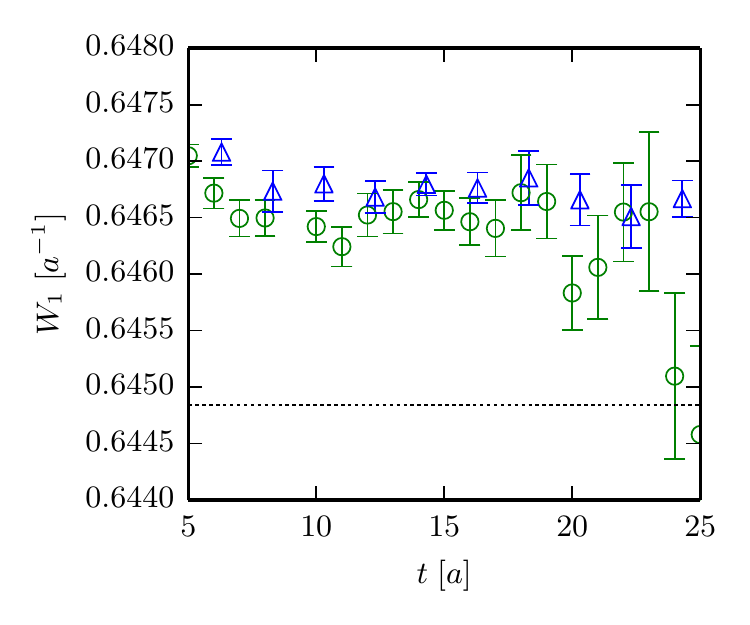}
        \end{subfigure}

        \vspace*{-.3in}
        \hspace*{-.3in}
        \begin{subfigure}[b]{0.47\textwidth}
                \includegraphics{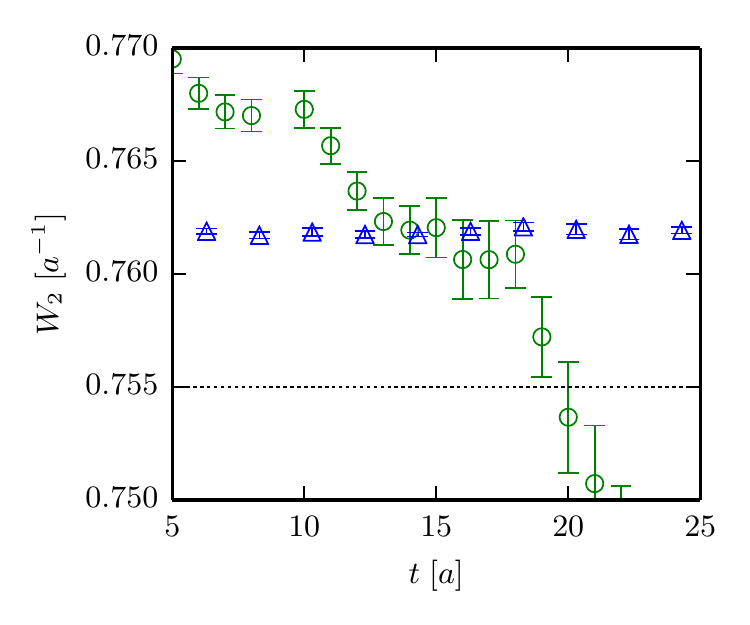}
        \end{subfigure}%
        \quad
        \begin{subfigure}[b]{0.47\textwidth}
                \includegraphics{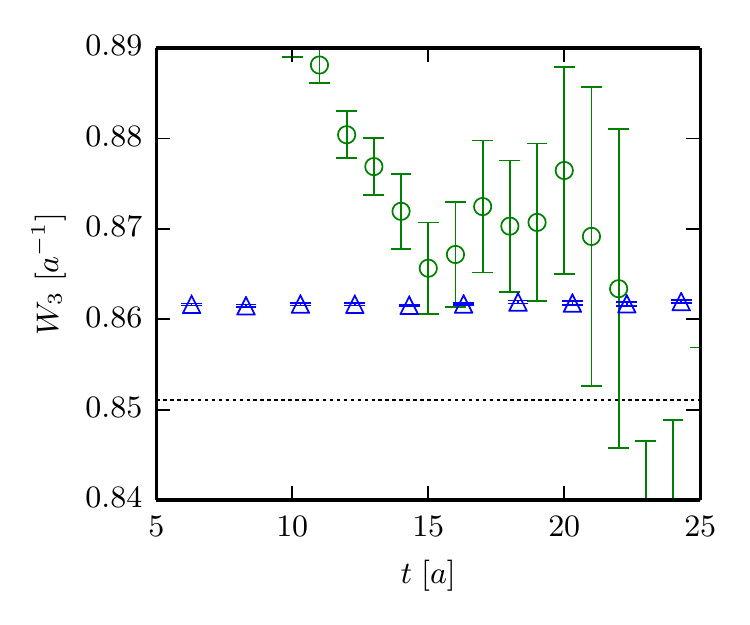}
        \end{subfigure}
        \caption{Comparison of the energies of the first 3 Hamiltonian eigenstates as extracted from the variational method (circles) and the potential method (triangles).}\label{fig:comp_energ}
\end{figure}

Given the previous observations, we extract the first three eigenstate energies from the eigenvalues of the matrix $H$ with $U$ taken as $U(t=14)$.
In pratice, we compute the lowest eigenvalues of the matrix and select those for which the associated eigenvectors ($N_s^3$-dimensional vectors which can be seen as wave functions) have dominant projections in the $A_1^+$ representation.

To estimate the dependence on the choice of $t$, we repeat the calculations for other operators $U(t)$.
This is what is plotted in Fig.~\ref{fig:comp_energ}, together with the results from the variational method.
The cirles show $W_n^\textrm{eff}(t)$ computed from the variational method while the triangles represent the energies obtained from the potential method.
The dashed lines show the energies of the free sytem.
The top left panel shows the first three eigenstates energies while the other panels are close-ups around each of them.
The results are consistent between the two methods while the ones from the potential method are more stable with the time index and carry smaller statistical errors.

\section{Summary}

We have presented a new method to extract the energy eigenstates of two-particle systems from lattice simulations.
This ``potential method'' makes use of lattice wave functions instead of square correlation matrices.
It is compared numerically in the $I=2$ channel of the $\pi\pi$ system with the usual variational method.
The results are shown to be consistent between the two method.
However, the result from the potential are more stable with the time index and have small statistical errors.
Furthermore, one source operator allows to extract the energies of the first three eigenstates of the Hamiltonian similarly as the variational method with four source operators.
Such a choice is advantageous when the most expensive part of the computation grows with the number of source but not with the number of sinks (e.g. computation of source-to-all propagators).
For both methods, the finite size formula can then be used to recover the phase shifts of the system.


Numerical computations in this work were carried out on SR16000 at YITP in Kyoto University.
We are also grateful for the authors and maintainers of CPS++ \cite{cps}, of which a modified version was used for measurements done in this work.

\end{document}